\documentclass[vecphys]{svmult}

\usepackage{makeidx}         
\usepackage{graphicx}        
\usepackage{multicol}        
\usepackage[bottom]{footmisc}

\makeindex             

\begin{document}

\title*{Young Star Clusters in M31}

\author{Heather Morrison\inst{1} \and
Nelson Caldwell\inst{2} \and
Paul Harding\inst{1} \and
Jeff Kriessler\inst{1} \and
James A. Rose\inst{3} \and
Ricardo Schiavon\inst{4} }


\institute{Department of Astronomy, Case Western Reserve University,
  10900 Euclid Ave, Cleveland OH 44106 USA
\texttt{hlm5@case.edu,paul.harding@case.edu,jeffrey.kriessler@case.edu} \and
Smithsonian Astrophysical Observatory, Cambridge, MA 02138, USA
  \texttt{caldwell@cfa.harvard.edu} \and
Department of Physics and Astronomy, University of North Carolina,
  Chapel Hill, NC 27599, USA 
  \texttt{jim@physics.unc.edu} \and 
Department of Astronomy, University of Virginia, Charlottesville, VA
  22903-0818, USA
  \texttt{ripisc@virginia.edu}}

\authorrunning{Morrison et al.}
%
%
\maketitle

\begin{abstract}

In our study of M31's globular cluster system with MMT/Hectospec, we
have obtained high-quality spectra of 85 clusters with ages less than
1 Gyr. With the exception of Hubble V, the young cluster in NGC 205,
we find that these young clusters have kinematics and spatial
distribution consistent with membership in M31's young disk.
Preliminary estimates of the cluster masses and structural parameters,
using spectroscopically derived ages and HST imaging, confirms earlier
suggestions that M31 has clusters similar to the LMC's young populous
clusters.

\end{abstract}

\section{Introduction}
\label{sec:1}

In the Milky Way, there is a clear separation between open clusters
(which have diffuse structure, generally have low masses and ages, and belong to the disk)
and globular clusters (which have a more concentrated
structure, higher masses and ages, and where the majority belong to the
halo). Other Local Group galaxies, however, have more complex cluster
populations. For example, the LMC has ``populous blue'' clusters,
which are young, structurally resemble the Milky Way globulars, and
have masses which overlap the globular cluster range. It has been
suggested \cite{kennicutt88} that these populous blue clusters are
found in late-type galaxies only; more recently these populous blue
clusters have also been compared to the ``super star clusters'' formed
in galaxies with very high star formation rates \cite{larsen}. Both are of
interest in understanding globular cluster formation.

What of M31's clusters? Remarkably, it is only recently that detailed
constraints on the cluster populations have been obtained,
particularly for clusters projected on the inner disk and bulge. HST
imaging and multi-fiber spectroscopy in particular have played an
important role here.  This paper describes 85 M31 clusters, originally
classified as globulars, which belong to the disk and have properties
in common with both the Milky Way open clusters and the LMC populous
blue clusters.

The existence of young clusters in M31 was noted in work focused
both on its globular clusters (eg
\cite{vdb},\cite{barmby},\cite{beasley},\cite{burstein},\cite{fusipecci})
and on its open clusters (eg
\cite{hodge79},\cite{hodge87},\cite{williams01}). In general, authors
have associated these young clusters with M31's disk, although
\cite{burstein} invoke an accretion of an LMC-sized galaxy by
M31. Observations are challenging for clusters projected on M31's
disk: many of the early velocities had large errors, there were issues
with background subtraction. Here we discuss high-quality
spectroscopic measurements of kinematics and ages for the young
clusters, supplemented with HST imaging to delineate
the structural, spatial and kinematical properties of these young
clusters.

\section{Observations and Discussion}

We obtained data in observing runs on the MMT using Hectospec in 2004
to 2006. We now have high-quality spectral observations of over
350 confirmed clusters in M31. We used the 270 gpm grating, which gave
spectral coverage from 3650--9200\AA with a resolution of $\sim$5\AA.
In order to sky subtract even in the bright central regions of M31's
disk and bulge, we obtained a number of offset sky exposures to
supplement the normal sky fibers.

Using the models presented in \cite{leonardi} for Hdelta/Fe4045 and
CaII indices to measure ages, we have identified 85 clusters with ages
less than 1 Gyr. We have ACS imaging of a number of these clusters,
and find that the ages obtained directly from the CMDs of the clusters
agree quite well with the spectroscopic ages.

Inspection of a Spitzer/MIPS 24$\mu$ image of M31 \cite{gordon}
showed that the spatial distribution
of the young clusters is  well correlated with the star-forming
regions in M31, with the majority associated with the 10 kpc ``ring of
fire''.  The kinematics of the young clusters bear out this disk
association. The offset sky fibers provide us with an accurate map of
the mean disk rotation throughout these inner regions, and the young
cluster velocities follow the mean disk velocities closely.

Given the close similarity in spatial distribution and kinematics
between the young clusters and other young disk objects, we now
investigate both the structural properties and mass distribution of
these young clusters.

\cite{cohen} highlighted the heterogeneous quality of the M31 cluster
catalogs when they found, using LGSAO on Keck II, that four of the six
young clusters observed were in fact asterisms: accidental
superpositions of stars which had appeared resolved on the original
source material. High spatial resolution imaging can both check for
asterisms and also explore the clusters' spatial structure: is their
concentration low, like typical Milky Way open clusters, or high, like
globular clusters?
There are ACS or WFPC2 images available for 19 of the young clusters.
Two of these show no evidence of an underlying cluster, but the
remaining 17 are clearly not asterisms. Interestingly, while 12 show
the typical low-concentration structure typical of Milky Way open
clusters, five of them are quite centrally concentrated, resembling
the LMC populous clusters.

We have made rough estimates of the mass of these M31 young disk
clusters, as follows. We used the V-band images from the Local Group
Survey \cite{massey} to derive new V magnitudes for all these clusters
on a consistent and accurate system. We calculated M/L ratios from the
spectroscopic age and Z measurements using the formalism in
\cite{leonardi}. We have corrected
for foreground reddening and a little reddening
from M31's disk (E(B-V)=0.2 in total).  We expect that it will be much higher in
some cases (some clusters have E(B-V) as high as 1.3
\cite{barmby}, which will lead to a mass estimate which is too low by
more than an order of magnitude). Thus our mass estimates are lower
limits, and in some cases the true masses will be significantly
higher. The mass histogram for the young clusters is shown in Fig. 1.

\begin{figure}
\centering
\includegraphics[height=9.5cm]{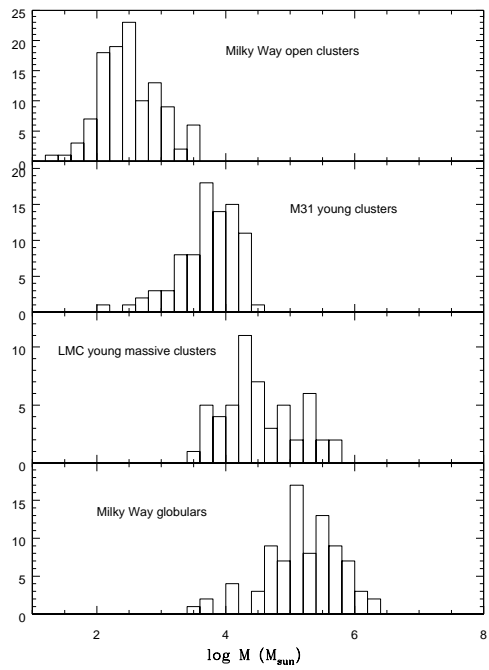}
\caption{Mass histograms, from top to bottom, of Milky Way open clusters, M31 young clusters, LMC massive
  clusters and Milky Way globular
  clusters}
\label{fig:1}       
\end{figure}

We have also shown the mass distribution of Milky Way open
clusters within 600 pc of the Sun, from \cite{khar}, with mass
calculations by \cite{henny}. 
This catalog will not include the most massive clusters in the
Galaxy because of its relatively small sample size; for example, there
have been recent discoveries of more distant young clusters which may
have masses as high as $10^5 M_\odot$ (eg \cite{clark}).  The Milky Way
globular and LMC young massive cluster histograms are shown in
the bottom two panels (from \cite{dean}).  

There is a trend in cluster mass,
with the Milky Way open clusters having the lowest median mass, the
Milky Way globulars the highest, and the LMC young massive clusters
and the M31 young clusters in between. Our work confirms earlier
claims that young populous clusters exist in M31 \cite{williams01,
burstein, fusipecci}: these are not restricted to late-type galaxies.

\section{Summary}

We have high-quality spectra taken with MMT/Hectospec for 85 star
clusters in M31 with ages less than 1 Gyr. The clusters have spatial
and kinematical properties consistent with formation in the
star-forming disk of M31, and structural parameters (from HST imaging)
ranging from the low concentrations typical of Milky Way open clusters
to the higher concentrations of Milky Way globulars and LMC populous
blue clusters.
Initial estimates of their masses using spectroscopic ages and new
photometry from the Local Group Survey show that some young
clusters have masses similar to the populous blue clusters in the LMC:
such clusters are not restricted only to late-type galaxies or to
galaxies with a very high star formation rate.

%
%
%

%
%

%
%



\printindex
\end{document}